\documentclass[11pt,a4paper]{article}
\usepackage[utf8]{inputenc}
\usepackage[T1]{fontenc}
\usepackage[margin=2.5cm]{geometry}
\usepackage{graphicx}
\usepackage{float}
\usepackage{hyperref}
\usepackage{setspace}
\usepackage{titlesec}
\usepackage{enumitem}
\usepackage{caption}
\usepackage{amsmath}
\usepackage{parskip}
\usepackage{ragged2e}
\usepackage{subcaption}

\begin{document}

\newcommand{\Bme}{\texttt{BIDSme }} 

\begin{flushleft}
    {\LARGE \textbf{Containerizing  \Bme: A Reproducible Tool for BIDS Conversion}}\\[0.5cm]

    \normalsize
    \textbf{Bradley Spitz\textsuperscript{1,2}, Antoine Jacquemin\textsuperscript{2}, Nikita Beliy\textsuperscript{2}, Christophe Phillips\textsuperscript{2}} \\[0.2cm]

    \textsuperscript{1}Télécom Physique Strasbourg, Université de Strasbourg, France\\
    \textsuperscript{2}GIGA -- CRC Human Imaging, University of Liège, Liège, Belgium\\[1cm]
\end{flushleft}

\section*{Abstract}
\addcontentsline{toc}{section}{Abstract}
The "Brain Imaging Data Structure" (BIDS)  has become a widely adopted standard for organizing and sharing neuroimaging datasets of various modalities. However, converting raw brain imaging data into BIDS framework remains a complex and time-consuming task. \Bme is a semi-automated tool developed to streamline this conversion process, but until recently, it lacked the portability and accessibility needed for widespread adoption. This paper presents the containerization of \Bme using Docker and Docker Compose, improving usability, reproducibility, and integration into existing platforms like Neurodesk. It also details the design choices, iterative refinements, and validation process that led to a flexible, lightweight, and user-friendly containerized application.

\newpage

\section{Introduction}
\addcontentsline{toc}{section}{Introduction}

\begin{justify}
    The Brain Imaging Data Structure (BIDS) scheme \cite{gorgolewski_brain_2016} has emerged as a community-driven standard for organizing and sharing neuroimaging data, helping reproducibility and interoperability across studies and software tools \cite{BIDShisto}. However, transforming "raw" \footnote{directly coming the acquisition system, or with minimal processing such DICOM-to-NIfTI conversion.} neuroimaging data 
    into a valid BIDS format remains a complex, error-prone, and often manual process. This stems from variability in possible acquisition protocols and outlier cases in the data?
    
    \Bme is an open-source Python application 
    developed at the GIGA-CRC Human Imaging research unit to facilitate this "BIDS-ification" of neuroimaging data \cite{bidsme_tool}. It provides a configurable, user-guided workflow for renaming, restructuring, and enriching datasets with metadata, allowing flexibility beyond rigid BIDS-ification tools \cite{beliy_bidsme_2023}.

    \begin{figure}[H]
        \centering
        \includegraphics[width=12cm]{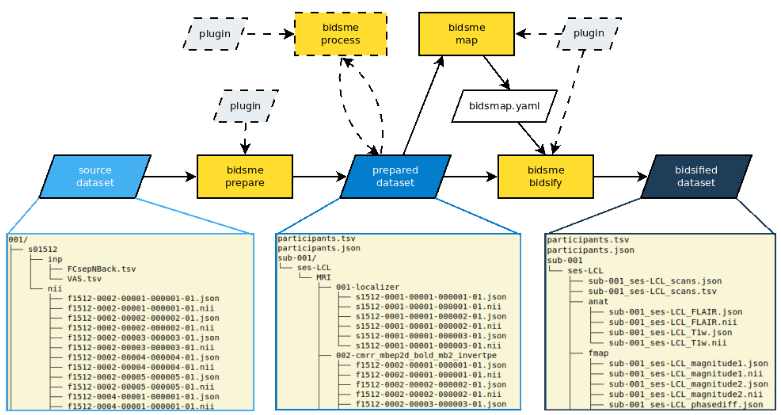}
        \caption{General workflow of \Bme: from raw neuroimaging data to a BIDS-compliant dataset.}
        \label{fig:bidsme_workflow}
    \end{figure}
    
    Despite its potential, \Bme was not initially designed for broad distribution or easy deployment. It required local installation with multiple dependencies and lacked support for standardized, reproducible environments. This limited its accessibility for end-users, particularly those without technical expertise in Python or environment management.
    
    To overcome these limitations, we propose here a containerized version of \Bme using Docker \cite{merkel2014docker}, along with a simplified interface and integration into platforms like Neurodesk \cite{Renton2024}. This paper outlines the containerization process, design choices, usability improvements, and validation strategies, aiming to make \Bme a robust and portable tool for the neuroimaging community.
\end{justify}

\section{Software description}
\addcontentsline{toc}{section}{Software Description}
\begin{justify}

    \subsection{Overview of \Bme}
    \Bme is an open-source Python tool developed to facilitate the conversion of raw neuroimaging datasets into the Brain Imaging Data Structure (BIDS) format \cite{gorgolewski_brain_2016}. Unlike fully automated converters, \Bme offers a semi-automated, user-guided approach, allowing flexibility to handle complex or heterogeneous datasets typically found in research settings.
    
    Its modular design enables users to perform key steps such as renaming raw files, reorganizing directory structures, and mapping metadata fields, thereby reducing manual effort and minimizing errors. The software supports a broad range of neuroimaging modalities and formats. To support new and experienced users, a tutorial and example datasets are also provided \cite{bidsme_tuto,bidsme_example}.
    
    \subsection{Key functionalities}
    The core functionalities of \Bme are implemented as four main commands:
    
    \begin{itemize}
        \item \texttt{prepare}: Organizes and preprocesses raw data by renaming files and arranging them into an intermediate directory structure that facilitates subsequent checks and BIDSification.
        \item \texttt{map}: Maps and transfers relevant metadata and filenames from the prepared data to the final BIDS-compliant structure.
        \item \texttt{bidsify}: Finalizes the conversion by applying BIDS-compliant naming conventions.
        \item \texttt{process}: Processes the bidsified dataset prior to final BIDSification. This command can be used to produce derivatives, perform data conversion, or execute anonymization while maintaining the advantage of recording identification through the use of \texttt{bidsmap.yaml}. Essentially, it performs all processing steps similar to BIDSification but without executing the BIDSification itself.
    \end{itemize}

    These operations can be executed sequentially or individually, providing users with fine control over the conversion workflow.

    \subsection{Limitations of the original implementation}
    
    Initially, \Bme required users to manually install Python dependencies and set up the environment locally, which posed challenges in terms of portability, reproducibility, and ease of use. Dependency conflicts, differences in system configurations, and the need for command-line proficiency limited its accessibility to less technical users.
    
    The containerization effort presented here addresses these issues by packaging \Bme and all its dependencies into a portable and reproducible Docker image, enhancing usability through dual Command Line Interface (CLI) and JupyterLab modes \cite{Kluyver2016jupyter}, and facilitating integration into broader neuroimaging platforms.
    
\end{justify}

\section{Containerization approach}
\addcontentsline{toc}{section}{Containerization approach}
\begin{justify}
    The containerization of \Bme is guided by the motivation of ensuring its portability and reproducibility, by notably encapsulating all of its dependencies and environment configurations, finally leading to an easier deployment across diverse systems. 

    \subsection{Base image selection}
        The process begins with the creation of our Dockerfile. The Dockerfile is a script defining how to assemble the container image, specifying the base environment, dependencies, code, configuration, and runtime behavior.
        
        We first start the Dockerfile by choosing a base image. For \texttt{BIDSme}, a lightweight Python image such as \textbf{python:3.9-slim-bullseye} was chosen to provide a balance between minimal image size and broad compatibility.
        Using a slim variant minimizes the overall container size by excluding unnecessary packages and utilities, which is advantageous for efficient storage and faster image transfers.

    \subsection{Dependency management}

        A critical aspect of containerization is dependency management. \Bme depends on both system-level libraries and a set of Python packages, which are clearly defined in its \textbf{setup.py} file.
        
        At the system level, \Bme requires essential neuroimaging tools, as well as libraries for GUI support and compression, among others. On the Python side, the core packages include \textbf{pandas} \cite{mckinney2010data}, \textbf{pyparsing} \cite{pyparsing}, \textbf{bidsschematools}, and several others.         
        \Bme also offers optional extras to tailor functionality according to user needs, such as \textbf{nibabel} \cite{nibabel_release}, \textbf{pydicom} \cite{mason2011t}, \textbf{dcm2niix} \cite{Li2016dcm2niix}, and \textbf{dicom\_parser} \cite{DCM-parser}, which are installed together to provide full feature support.
        
        To guarantee that all these dependencies are correctly installed within the container, the Dockerfile retrieves the \Bme source code and installs both the core packages and the extras in a single \texttt{pip install} command (e.g., \texttt{pip install ".[all]"}), ensuring that the container has all necessary components pre-installed for seamless execution.

\medskip

    \subsection{Multi-stage build strategy}
    
        To optimize the final image size and separate the build environment from the runtime environment, a multi-stage build approach is employed, as shown in figure \ref{fig:multistagev1}.
        
        The first stage ("build stage"), see figure \ref{fig:multistage1}, contains the base image selection, installs all system dependencies, clones the \Bme repository, and performs the installation of Python packages including optional extras and JupyterLab for interactive use. 
        The decision was made to directly clone the Git repository within the container build process. This approach provides full access to the latest version of the application without requiring any manual or redundant local copy prior to building the Docker image, thereby automating the containerization workflow.
        
        The second stage ("runtime stage"), see figure \ref{fig:multistage2}, starts from a fresh base image where only the minimal system dependencies are re-installed. It then copies the pre-installed Python environment from the "build stage", alongside necessary scripts, configuration and environment variables.
        This separation allows the final container to be cleaner and more secure by excluding unnecessary build tools and caches, while also preserving all required runtime functionalities.

        \begin{figure}[H]
                  \centering
                  \begin{subfigure}[b]{0.50\textwidth}
                    \centering
                    \includegraphics[width=\textwidth]{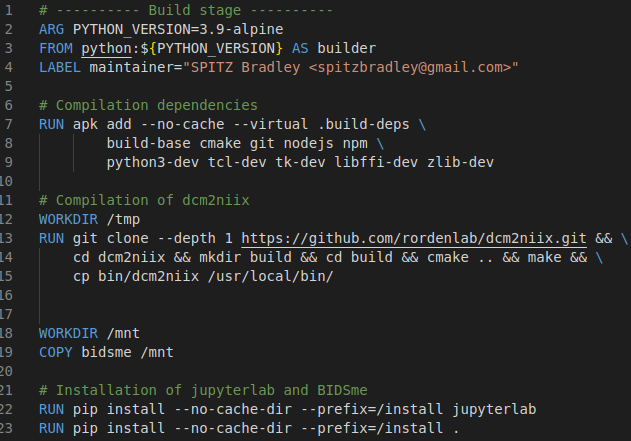}
                    \caption{Building stage}
                    \label{fig:multistage1}
                  \end{subfigure}
                  \hfill
                  \begin{subfigure}[b]{0.45\textwidth}
                    \centering
                    \includegraphics[width=\textwidth]{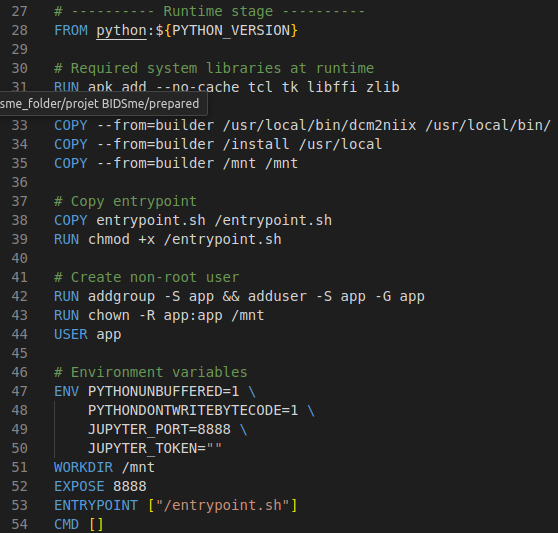}
                    \caption{Runtime stage}
                    \label{fig:multistage2}
                  \end{subfigure}
                  \caption{The multi-stage Dockerfile to optimize the image building process}
                  \label{fig:multistagev1}
                \end{figure}

    \subsection{Directory structure and mounting strategy}
        To support \texttt{BIDSme}’s file processing workflow and promote clarity across executions, a well-structured and consistent directory layout was established inside the container. All user-facing operations are carried out under a single shared workspace, typically mounted at \texttt{/mnt}, where input, output, and configuration files are organized.
        
        The default expected structure includes the following subdirectories:
        \begin{itemize}
            \item \texttt{/mnt/rawdata} - contains the original neuroimaging files to be processed.
            \item \texttt{/mnt/prepared} - serves as the output directory for pre-processed data, ready for BIDSification.
            \item \texttt{/mnt/bidsified} - stores the final BIDS-compliant dataset.
            \item \texttt{/mnt/configuration} - holds configuration files and plugins such as \texttt{bidsmap.yaml} for flexible mapping logic.
        \end{itemize}
        
        This organization improves clarity, promotes automation, and aligns closely with \texttt{BIDSme}’s internal logic, see figure \ref{fig:bidsme_workflow}.
        To make this layout persistent beyond the life-cycle of the container, Docker volumes are used to bind host directories to the corresponding paths inside the container. Thus processed data and user configurations are not lost when the container stops or is removed.

    \subsection{Management of the user environment and permissions}

        To ensure compatibility with host file systems and avoid permission issues --- especially when writing output data from within the container --- a non-root user is created during the runtime stage of the Dockerfile. The user ID and group ID can be passed as build arguments to align with the host user's permissions. This helps prevent issues related to inaccessible files after processing and makes the container usable across different systems and setups.

    \subsection{Entry point and multi-interface support}
        To improve usability and offer more flexibility, a custom \texttt{entrypoint.sh} script was added to the container. This script detects the mode of execution requested by the user and automatically launches the appropriate interface: \textbf{Command Line Interface (CLI)}, \textbf{Interactive Shell}, or \textbf{JupyterLab}. This unified entry point simplifies container usage across different workflows and user preferences.

        In addition, a preconfigured Jupyter notebook is included within the container. When JupyterLab is launched, this notebook serves as a starting point.        
        Details on how each interface can be accessed and used are provided in Section \ref{sec:Usability}. 
        Overall, these various steps allow the application to be fully containerized and operational, resulting in a final image size of approximately 1GB.
\end{justify}

\section{Testing and validation}
\addcontentsline{toc}{section}{Testing and validation}
\begin{justify}
Once the container was built, because of the multitude of configurations and modes of functioning, some testing and validation was absolutely required.
 
    \subsection{Functional verification}
     Rigorous functional testing was performed to ensure that \Bme's core features operated correctly within the Dockerized environment. We relied on the example data \cite{bidsme_example} and tutorial \cite{bidsme_tuto} provided with \Bme.  The validation covered the full pipeline, including the commands:
    \begin{itemize}
        \item \texttt{prepare}: to restructure and rename raw data before BIDSifying them.
    
        \item \texttt{map}: to transform the prepared data into a BIDS-compatible structure using \texttt{bidsmap.yaml}.
    
        \item \texttt{bidsify} and \texttt{process}: for full BIDSification or additional post-processing steps.
    \end{itemize}

    These commands were executed both through the command-line interface and via Jupyter notebooks. The correctness of the output directories (\texttt{prepared/}, \texttt{bidsified/}) and the absence of runtime errors were used as criteria for validation.

    \subsection{Interface testing}
    To ensure consistency and reliability across usage modes, each interface --- CLI, interactive shell, and JupyterLab --- was tested independently. The CLI was validated using direct \texttt{docker run} commands, confirming that input/output volumes were properly mounted and outputs were written as expected. The JupyterLab interface was tested to ensure that notebooks could be executed end-to-end, and that preloaded configurations and paths were properly set up.
    %

    \subsection{Data consistency and output validation}
    To confirm that the container did not alter the logic or quality of the original \Bme tool: an example dataset \cite{bidsme_example} were processed through both local, i.e. non-containerized, and Dockerized versions of \texttt{BIDSme}. The resulting outputs were compared for structural consistency and file integrity using automated directory checksums and visual inspections of BIDS-compliant folders. No differences were detected between the two outputs, thus validating the correctness of the containerized version.
    
%
%

    \subsection{Reproducibility assurance}
    To test reproducibility, the same dataset was processed multiple times under identical container configurations across different machines, running Windows or Linux OS. The outputs were identical in both structure and content, demonstrating that the container provides a reproducible environment across platforms. This process also helped identify potential permission issues arising from Docker’s default behavior, which often assigns \texttt{root} ownership to files and directories created within containers.

\end{justify}

\section{Usability and distribution enhancements}
\label{sec:Usability}
\addcontentsline{toc}{section}{Usability and dDistribution enhancements}
\begin{justify}

    To improve the accessibility, maintainability, and practical deployment of the containerized version of \texttt{BIDSme}, several usability-focused tools and structures were developed and published alongside the Dockerfile.

    \subsection{Environment management via Docker Compose}
    To streamline usage and avoid the need to repeatedly type long and error-prone \texttt{docker run} commands with multiple volume mounts and environment variables, a Docker Compose-based setup was introduced. Two separate \texttt{docker-compose.yml} files were created:

    \begin{itemize}
        \item A \textbf{development version}, designed to launch the container in JupyterLab mode with persistent notebooks and environment variables useful for debugging and testing.
        \item A \textbf{production version}, meant for regular use and deployment, also with a persistent production notebook.
    \end{itemize}

    Depending on the options provided after the \texttt{lab} argument, the \texttt{entrypoint.sh} script and the \texttt{init\_bidsme\_lab.py} file dynamically initialize default paths by assigning them to well-defined variables. This mechanism streamlines the workflow by simplifying file access and reducing manual configuration within the JupyterLab environment.

    Moreover, the Docker Compose files automatically mount the required volume structure under \texttt{/mnt} and set key environment variables such as user ID, group ID, and/or JupyterLab port. This approach enhances reproducibility and minimizes the risk of user error during setup.

    \subsection{Helper scripts and build wrappers}
    To further simplify the image management, wrapper scripts were implemented for both building and running the container. For example:
    \begin{itemize}
        \item A \texttt{build.sh} script condenses the Docker build process, handling arguments such as the desired \Bme version. The script also fetches and tags the correct Git commit from the \Bme repository, ensuring the containerized version is tightly coupled to the underlying software version.
        
        \item \texttt{bidsme\_prepare.sh} is a convenience script that automates the initialization of a dataset in a single step. It ensures correct directory creation, mounts, and command execution to perform the initial \texttt{prepare} phase in a quick and reliable way.
    \end{itemize}
    
    These scripts ease the task of contributors and users to generate consistent, reproducible containers while maintaining compatibility with upstream changes.
    
    \subsection{Multi-interface access modes}
    
    To accommodate different usage profiles and levels of technical expertise, the container supports three primary modes of operation, all unified under a single \texttt{entrypoint.sh}:
    \begin{itemize}
        \item the \textbf{CLI mode} allows direct execution of \Bme commands such as \texttt{prepare}, \texttt{map}, or \texttt{bidsify} via Compose or Docker CLI.
        \item the \textbf{interactive shell mode} launches a bash shell into the container to explore its environment, test commands manually, or inspect intermediate files.
        \item the \textbf{JupyterLab mode} opens a web-based notebook interface with preloaded paths, allowing easier exploration, configuration, and execution for users less comfortable with the command line.
    \end{itemize}
    
    Depending on how the container is started (via CLI or Compose), the appropriate mode is automatically triggered by the \texttt{entrypoint.sh} logic, see Figure \ref{fig:entrypoint_logic}. A dedicated configuration notebook is also embedded in the container to assist users launching via JupyterLab.

    For example, the JupyterLab interface in production mode can be launched using a full Docker command:

    \begin{verbatim}
        docker run -p 8888:8888 \
          -v $(pwd)/rawdata_prod:/mnt/rawdata \
          -v $(pwd)/prepared_prod:/mnt/prepared \
          -v $(pwd)/bidsified_prod:/mnt/bidsified \
          -v $(pwd)/configuration_prod:/mnt/configuration \
          bidsme:<version> lab prod
    \end{verbatim}

    Alternatively, the same setup can be started using the predefined production Compose file:
    
    \begin{verbatim}
        docker compose -f docker-compose.prod.yml run --service-ports bidsme lab
    \end{verbatim}
    
    Each method provides consistent behavior, with automatic initialization of environment variables and preloaded notebook paths for a smooth user experience.
    Notably, Docker Compose proves especially useful by reducing a long and error-prone multi-line Docker command to a single, concise instruction. 

    \begin{figure}[H]
        \centering
        \includegraphics[width=12cm]{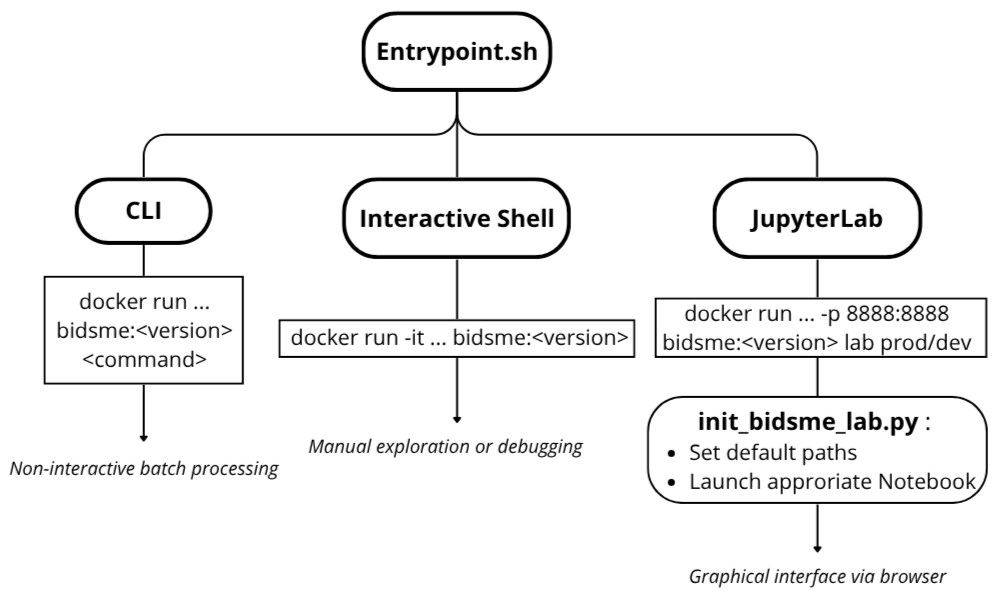}
        \caption{Dockerfile's entry point logic}
        \label{fig:entrypoint_logic}
    \end{figure}
    
    \subsection{Public distribution via GitHub}
    The full containerization project is hosted on GitHub \cite{SPITZ_bidsme_containerisationcitationcff_2025}, providing open access to the Dockerfile, helper scripts, and example configurations. This ensures transparency, version control, and easy collaboration across users and contributors. The repository also includes documentation to guide new users through installation, setup, and execution of the container.

\end{justify}

\section{Integration into Neurodesk}

Neurodesk is an open-source, containerized computing environment designed to simplify and standardize neuroimaging data analysis \cite{Renton2024}. Distributing \Bme through Neurodesk would ensure a broader distribution and potentially easier integration in various data handling pipelines.

    \subsection{Integration process}
    \begin{justify}
        
    \setlength{\parskip}{1.5em}
    \setstretch{1.1}
    The integration process was carried out by creating a YAML configuration file that describes the \Bme application's specifications, metadata, and build instructions. This YAML file acts as the blueprint used by Neurodesk's automated build system:
    
    \begin{verbatim}
    name: bidsme
    version: 1.9.3
    architectures:
      - x86_64
    categories:
      - data organisation
      - bids app
    build:
      kind: neurodocker
      base-image: bradley987/bidsme:1.9.3
      pkg-manager: apt
    \end{verbatim}
    
    Unlike other Neurodesk applications that are built from a minimal Ubuntu base image, the integration of \Bme relied on an existing Docker image that had already been containerized and optimized. This approach ensured a reliable and reproducible environment while reducing development overhead. The YAML file additionally specifies metadata such as documentation links, licensing, and an automated test script to verify that the command-line interface is available:
    
    \begin{verbatim}
    - test:
        name: testScript
        script: |-
          #!/bin/bash
          bidsme --help
    \end{verbatim}
    
    The integration leverages Neurodesk’s automated build pipeline, which processes the YAML file and generates the corresponding Singularity container image. The workflow includes GitHub submission, validation, container build, optimization, and final deployment into the Neurodesk registry. Since the integration is based on the preconfigured \texttt{bradley987/bidsme:1.9.3} Docker image, the build remains lightweight and easily updatable.
    \end{justify}
    
    \subsection{Usage and Validation}
    \begin{justify}

    \setlength{\parskip}{1.5em}
    \setstretch{1.1}
    Within the Neurodesk environment, \Bme is currently accessible through the module loading mechanism. This ensures that the tool is properly initialized within the containerized environment before use:
    
    \begin{itemize}
        \item \textbf{Module Loading}:
        \begin{verbatim}
        import module
        await module.load('bidsme/1.9.3')
        \end{verbatim}
        
        \item \textbf{Command Execution} (in JupyterLab or terminal):
        \begin{verbatim}
        !bidsme --help
        !bidsme prepare /path/to/data /path/to/output <options>
        \end{verbatim}
    \end{itemize}
    
    At the current stage, \Bme cannot be directly imported as a Python package within Neurodesk native Python environment. Access is therefore limited to its CLI once the module is loaded, or through a JupyterLab notebook if launched afterwards.  
    
    Validation involved checking installation correctness, core functionality of dataset preparation and BIDSification, module-loading behavior, and user experience in both JupyterLab and terminal contexts. These tests confirmed that the tool runs reliably inside the Neurodesk environment. 
     \end{justify}
     
    \subsection{Benefits and Impact}
    \begin{justify}

    \setlength{\parskip}{1.5em}
    \setstretch{1.1}
    The successful integration of \Bme into Neurodesk delivers multiple benefits for the neuroimaging research community:
    
    \begin{itemize}
        \item \textbf{Accessibility}: \Bme can be accessed immediately without requiring manual installation
        \item \textbf{Reproducibility}: A consistent and controlled containerized environment ensures reproducible results
        \item \textbf{Collaboration}: Researchers share the same standardized BIDS conversion tools
        \item \textbf{Education}: Provides newcomers with preconfigured access to \texttt{BIDSme}, lowering the entry barrier
        \item \textbf{Maintainability}: Centralized updates via Neurodesk simplify long-term maintenance
    \end{itemize}
    
    Overall, this integration significantly improves the accessibility of \Bme and facilitates the adoption of BIDS standards across diverse research environments.
     \end{justify}

\section{Conclusions}
\addcontentsline{toc}{section}{Conclusions}
\begin{justify}
     The actual containerized version of \Bme offers significant improvements in reproducibility and ease of deployment. By encapsulating all dependencies and configuration, it eliminates the variability often encountered when neuroimaging pipelines are run on different systems. Providing several ways to use the containerized app gives users the freedom to perform processing according to their preferences and needs, 
     while reducing the tediousness of this kind of work.

    Despite the advantages provided by the containerized setup, the final image remains relatively large (approximately 1 GB), which may pose challenges for environments with limited storage or bandwidth.
    Additionally, using the container without Docker Compose can be bothersome, as it requires manually specifying long and complex volume-mounting commands.
    Additionally, although containerization improves portability, a basic familiarity with Docker is still necessary for effective usage, which may represent a barrier for users without prior experience in container technologies.

    \Bme has already been successfully integrated into the Neurodesk ecosystem as a containerized application. This ensures accessibility and reproducibility across platforms, in line with Neurodesk’s modular container-based approach. 
    However, since \Bme is a pure Python tool, keeping it inside a dedicated container is not strictly necessary. A more sustainable approach would be to make \Bme directly available within Neurodesk’s shared Python environment (via a virtual environment installation). This would reduce maintenance overhead, lower image size, and simplify updates, while still providing users with the same level of accessibility.  

    Future work will therefore focus on discussing with the Neurodesk team whether such an installation strategy could be adopted, ensuring that \Bme remains fully integrated while avoiding the unnecessary burden of maintaining a standalone container.   

\section*{Acknowledgments}
B. Spitz received financial support from the Erasmus+ program of the European Union and from the Région Grand Est (France), through its international mobility support program, to support his research internship at the GIGA -- CRC Human Imaging, University of Liège, Belgium. C. Phillips is supported by the F.R.S.-FNRS., Belgium. The original development of \Bme, as well as N. Beliy at the time, were supported by an Excellence of Science grant (\#30446199, 2017, ``MEMODYN, The journey of a memory: dynamics of learning and consolidation in maturation and ageing'') from the F.R.S.-FNRS and the FWO, Belgium. 
   
\end{justify}

\bibliographystyle{IEEEtran} 
\bibliography{bibliographie} 
\end{document}